\documentclass[conference]{IEEEtran}
\IEEEoverridecommandlockouts
\usepackage{cite}
\usepackage{amsmath,amssymb,amsfonts}
\usepackage{algorithmic}
\usepackage{graphicx}
\usepackage{textcomp}
\usepackage{pifont}% http://ctan.org/pkg/pifont
\newcommand{\xmark}{\ding{55}}%
\usepackage{xcolor}
\def\BibTeX{{\rm B\kern-.05em{\sc i\kern-.025em b}\kern-.08em
    T\kern-.1667em\lower.7ex\hbox{E}\kern-.125emX}}
\begin{document}

\title{Panoptic segmentation with highly imbalanced semantic labels\\

\thanks{$^*$ equal contribution}%Identify applicable funding agency here. If none, delete this.
}

\author{\IEEEauthorblockN{Josef Lorenz Rumberger$^{*1,2,6}$, Elias Baumann$^{*5}$, Peter Hirsch$^{*1,2}$,\\ Andrew Janowczyk$^{3,4}$, Inti Zlobec$^{5}$, Dagmar Kainmueller$^{1,2}$}
\IEEEauthorblockA{
\textit{$^1$ Max-Delbrueck-Center for Molecular Medicine in the Helmholtz Association (MDC), Berlin, Germany}\\
\textit{$^2$ Humboldt-Universität zu Berlin, Faculty of Mathematics and Natural Sciences, Berlin, Germany}\\
\textit{$^3$ Case Western Reserve University, Department of Biomedical Engineering, USA}\\
\textit{$^4$ Lausanne University Hospital, Precision Oncology Center, Switzerland}\\
\textit{$^5$ University of Bern, Institute of Pathology, Bern, Switzerland}\\}
\textit{$^6$ Charité University Medicine, Berlin, Germany}\\
}
%\IEEEauthorblockA{\textit{Kainmueller Lab, Department of Pathology} \\
%\textit{Max-Delbrueck-Center for Molecular Medicine} \\
%\textit{in the Helmholtz Association (MDC)}\\
%\textit{Charité University Medicine, RadioEye}\\
%Berlin, Germany \\
%joseflorenz.rumberger@mdc-berlin.de}
%\and
%\IEEEauthorblockN{1\textsuperscript{st} Elias Baumann}
%\IEEEauthorblockA{\textit{Institute of Pathology} \\
%\textit{University of Bern}\\
%Bern, Switzerland \\
%elias.baumann@pathology.unibe.ch}
%\and
%\IEEEauthorblockN{1\textsuperscript{st} Peter Hirsch}
%\IEEEauthorblockA{\textit{Kainmueller Lab, Faculty of Mathematics and Natural Sciences} \\
%\textit{Max-Delbrueck-Center for Molecular Medicine} \\
%\textit{in the Helmholtz Association (MDC)}\\
%\textit{Humboldt University Berlin}\\
%Berlin, Germany \\
%peter.hirsch@mdc-berlin.de}
%\and
%\IEEEauthorblockN{2\textsuperscript{nd} Dagmar Kainmueller}
%\IEEEauthorblockA{\textit{Kainmueller Lab} \\
%\textit{Max-Delbrueck-Center for Molecular Medicine} \\
%\textit{in the Helmholtz Association (MDC)}\\
%\textit{Humboldt University Berlin}\\
%Berlin, Germany \\
%dagmar.kainmueller@mdc-berlin.de}
%}

\maketitle       

% TODO Rewrite this for final version?
\begin{abstract}
We describe here the panoptic segmentation method we devised for our participation in the CoNIC: Colon Nuclei Identification and Counting Challenge at ISBI 2022. Key features of our method are a weighted loss specifically engineered for semantic segmentation of highly imbalanced cell types, and a state-of-the art nuclei instance segmentation model, which we combine in a Hovernet-like architecture. 
\end{abstract}


\section{Introduction}

% this is just ideas for intro written down
The tumor microenvironment is considered an important factor for tumor progression, immune response and drug resistance~\cite{Balkwill2012TheTM}. Although assessment of the tumor microenvironment can be performed using standard hematoxylin and eosin (H\&E) whole slide images (WSI), the differentiation between exact cell types can remain challenging in some settings without further molecular analysis (e.g. immunohistochemistry). 
Automated methods for cell / nuclei detection, segmentation and classification could be used to infer vast amounts of information from just the H\&E WSI to be used for clinical outcome prediction and investigation of cellular interactions within the tumor microenvironment. 

Current state-of-the-art automated methods for nuclei detection and classification 
% can provide accurate predictions however they are prone to errors on 
are challenged by domain shifts, differentiating between rare cell types, and distinguishing cells in dense regions~\cite{Schmidt2018CellDW, Graham2019HoverNetSS}.
%(cite stardist, hovernet).
The CoNIC: Colon Nuclei Identification and Counting Challenge~\cite{Graham2021CoNICCN} establishes a benchmark for respective comparative evaluation. The challenge consisted of two different tasks: \textit{Task 1: Nuclear segmentation and classification}, and \textit{Task 2: Prediction of cellular composition}.
The challenge dataset of H\&E stained tissue tiles comes with drastic semantic class imbalance: Among the six cell types described in~\cite{Graham2021LizardAL}, neutrophils and eosinophils each make up less than 1\% of the cells in the dataset. Furthermore, nucleus size varies considerably across cell types, while at the same time some cell types, though frequent, 
%epithelial cells, especially neoplastic ones, https://www.overleaf.com/project/6219fd47e1a6a1bbd9fa5b38
also exhibit high intra-class variation in shape (see Section~\ref{sec:data}). % and span on average more pixels than the immune cell subtypes.

Our proposed approach adopts ideas for importance sampling and loss weighting from the recent work of~\cite{Araslanov2021SelfsupervisedAC} to simultaneously handle class imbalance that stems from (1) cell types that rarely appear at all in an image, as well as (2) cell types that are small and thus occupy relatively few pixels in an image. We complement these ideas with a state-of-the-art nuclei instance segmentation model adapted from~\cite{Hirsch2020AnAT}. % For this, we adopt the view from~\cite{Araslanov2021SelfsupervisedAC} and distinguish between classes that have low image-level and low pixel-level frequency and treat both problems individually.

\section{Methodology}

%Elias: message me about modifying this figure
\begin{figure}[!h]
    \centering
    \includegraphics[width=2.5in]{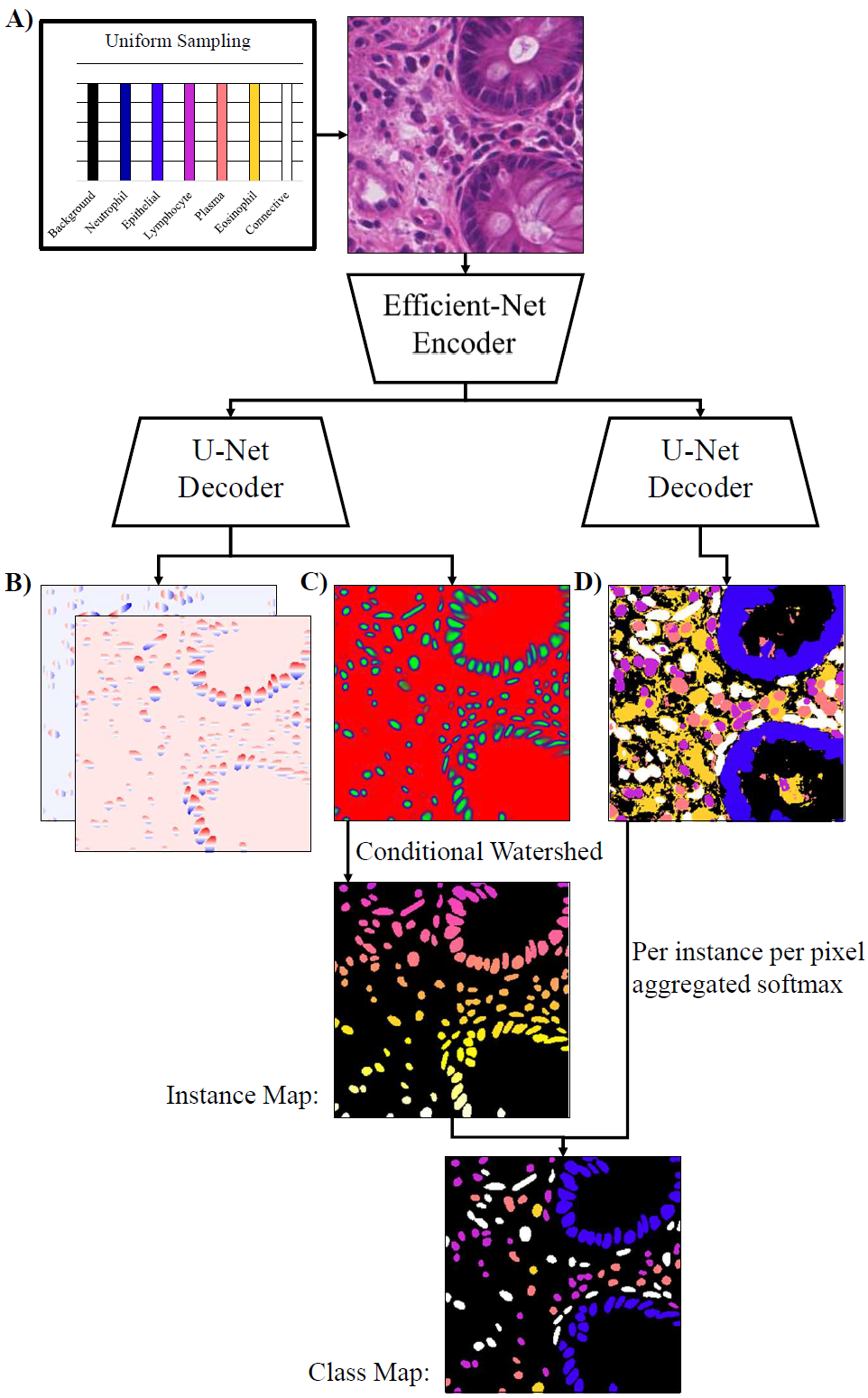}%{model_overview.pdf}
    \caption{Overview over the proposed model architecture: A) Pixel wise importance sampling during training, B) Center-point vector auxiliary task during training, C) Three label output (Background, Center and Boundary), D) Separate semantic class prediction output.}
    \label{fig:overview}
\end{figure}

\subsection{Importance sampling}
To treat image-level class imbalance, i.e., to cope with the fact that some classes rarely appear in an image at all, we adapt the training image sampling method presented in~\cite{Araslanov2021SelfsupervisedAC}: 
% We sample training images such that the class distribution in the sampled training data is uniform. To this end, 
We determine the probability for a pixel in training image $n$ to belong to class $c$ as
\begin{equation}
    X_{c,n} = \frac{1}{|h||w|} \sum\limits_{h,w} m_{c,n,h,w}
\end{equation}
where $m_{c,n,i,j}\in \{0,1\}$ denotes the semantic mask of class $c$ for training image $n$ at pixel index (h,w). 
We then calculate the probability for a training image to be drawn as 
\begin{equation}
    p_{n} = \frac{1}{|c|} \sum\limits_c \frac{X_{c,n}}{\sum_{n} X_{c,n}}
\end{equation} % changed from \sum_{m} X_{c,m}}
This has the effect that training images that do contain rare classes are preferably sampled. 
In each epoch, we sample with replacement a new subset from the training data using the sampling probabilities $p_{n}$. It should be noted that background is also considered a class.
% Inti asked: "I suppose the same images are resampled then? or is it that the method somehow "reduces" itself to the class with the lowest number of images... just a question " -> maybe we can rephrase the last two sentences so its clear that we sample a consistent number of tiles per epoch

%\textcolor{cyan}{Hey Dagmar, hier eine Erklärung.
% $\frac{X_{c,n}}{\sum_n X_{c,n}}$ denotes the share of pixels of class $c$ that are in sample $X_n$ relative to the total number of pixels of class $c$ in the whole dataset. E.g. there are $100$ pixels of class $6$ in sample $n$ and $1000$ pixels of class $6$ in the whole dataset, then $\frac{X_{6,n}}{\sum_n X_{6,n}} = \frac{100}{1000} = 0.1$. Then we sum for each sample over the classes $\sum\limits_c \frac{X_{c,n}}{\sum_n X_{c,n}}$ and get the share of pixels in sample $n$ relative to the total number of pixels of each individual class in the dataset. So we have a scalar for each sample $n$, if we add these scalars up we get $|c|$ because every share $\frac{X_{c,n}}{\sum_n X_{c,n}}$ of pixels of a class in a sample relative to the total pixels of that class in the dataset adds up to 1 when summed over the whole dataset, therefore we need to normalize via $\frac{1}{|c|}$ and that's how we arrive at the above formular $p_{n} = \frac{1}{|c|} \sum\limits_c \frac{X_{c,n}}{\sum_n X_{c,n}}$.
%}

\subsection{Weighted focal loss for semantic segmentation}
To cope with pixel-level class imbalance, i.e., the fact that some classes occupy fewer pixel per image than others, we adapt the focal loss from~\cite{Araslanov2021SelfsupervisedAC}. Based on the training labels we calculate an exponential moving average class prior 
\begin{equation}
    X^{t+1}_c = \gamma X^t_c + (1-\gamma) X_{c,n}.
\end{equation}
and update it with every training step. The loss is then calculated as weighted cross-entropy, with class weights $w_c = (1-X^{t+1}_c)^\rho$, with hyperparameter $\rho=3$ ~\cite{Araslanov2021SelfsupervisedAC}
and label smoothing parameter set to $0.05$. 

\subsection{Three-label model for instance segmentation}
We employ a three-label model (nucleus interior, nucleus boundary and background) for instance segmentation~\cite{Caicedo2019EvaluationOD} (c.f. Figure~\ref{fig:overview} C). We additionally regress nucleus center point vectors as auxiliary training task as in~\cite{Hirsch2020AnAT}. This has the effect that topological errors like false splits and false merges of nuclei constitute large losses, which is not necessarily the case for standard pixel-wise cross-entropy alone (c.f. Figure~\ref{fig:overview} B). These vectors are not used at test time. As training loss we employ the sum of the cross-entropy (classification) loss and the L2 (vector regression) loss, as in~\cite{Hirsch2020AnAT}.  

\subsection{Hovernet-like model architecture}
Our network architecture consists of the first 5 levels of an imagenet pre-trained EfficientNet encoder (B5, B7~\cite{Tan2019EfficientNetRM} or L2~\cite{Xie2020SelfTrainingWN}) with two U-Net~\cite{Ronneberger2015UNetCN} decoders with corresponding 5 levels and respective skip connections. We use $20\%$ dropout per layer in the decoder and encoder dropout as per EfficientNet definition. One of the decoders is optimized for the semantic loss, whereas the other is optimized for both instance segmentation losses. Compared to HoVer-Net \cite{Graham2019HoverNetSS}, we unify the two instance segmentation branches into one decoder with two loss functions and concatenate skip connections from encoder and the upsampling path.

\subsection{Training recipe}
We shuffle the dataset and split it into $4481$ training tiles and 500 validation tiles. For color augmentation during training, we apply color deconvolution to the RGB image to obtain an HED instensity image~\cite{Ruifrok2001QuantificationOH}. The image is then augmented by randomly adjusting each channel's contrast and transformed back to RGB. In addition, we apply RGB Color Jitter. Furthermore, gaussian blur and slight gaussian noise is applied. Spatial augmentations include random mirroring, translations, scaling, zooming, rotations, shearing and elastic transformations. We use AdamW optimizer~\cite{Loshchilov2019DecoupledWD} with $1e-4$ weight decay and cosine annealing learning rate scheduler. Models are trained for 600k training steps with batch size $4$ and the model with the highest $mPQ+$ on the validation set is picked as the best checkpoint.

\subsection{Test-time augmentations and inference}
During inference we apply random HED color augmentation and change staining intensities by $10\%$ as well as apply random flips and $90^{\circ}$ rotations. Furthermore we use dropout at test-time, since it has been shown that it improves segmentation performance when compared to dropout in validation mode~\cite{Gal2016DropoutAA,rumberger2020probabilistic}. We thus obtain $predictions$ from 16 forward passes, % 8 per model in the ensemble for \textit{Task 1}, 
and obtain our final prediction by pixel-wise averaging.

\subsection{Post-processing}
For each task, we draw per-class seeds from the predicted nucleus interior class and use watershed to delineate instances as described in~\cite{Hirsch2020AnAT}. Per-class foreground and seed thresholds for watershed are hyperparameters of our method, determined via grid search on the validation set, individually for each Challenge Task. To split distant false merges in the predictions, we apply the connected-components algorithm on each individual instance mask. 

We remove small as well as large instances, with class specific size thresholds determined via grid search on the validation set. Furthermore, we remove instances whose solidity (i.e., area divided by area of convex hull, see e.g.\ \cite{carpenter2006cellprofiler}) is below a cell type specific threshold, again determined via grid search. We fill any holes in each instance mask. 

For each instance mask we assign the cell type that maximizes the sum of the cell type's softmax scores over all pixels. 
% $T1: \in[20,45]px, T2: \in[18,30]px$
%  $T1,T2: \in[400,3000]px$
% $T1: \in[0.79,0.86], T2: \in[ , ]$

\section{Experimental Results}

\subsection{Data}\label{sec:data}

The challenge provides the previously published Lizard Dataset~\cite{Graham2021LizardAL} and required participants to exclusively make use of this dataset. The challenge organizers also provided a pre-tiled version of the same dataset which we used for all experiments. All following reported statistics are for the tile dataset. The dataset consists of $N=4981$ tiles ($256\times256px$ at $0.5\mu pp$). We randomly sampled 500 tiles for an internal validation on which all following results are reported unless specified otherwise. Classes are highly imbalanced, with eosinophil nuclei being only 0.1\% of the data and 0.68\% of all instances, whereas epithelial nuclei make up 49.5\% of all instances (see Table~\ref{tab:data_stats}). 
\begin{table}
\centering
\begin{tabular}{l|rrr}
Type & \% px. & \% Instances & mean inst. px. (std.) \\ \hline
background & 83.97 & 0.00 &   \\
epithelial cell & 10.31 & 49.50 & 119.29 +- 72.82 \\
connective-tissue cell & 3.08 & 22.10 & 79.86 +- 49.54 \\
lymphocyte & 1.85 & 21.22 & 50.06 +- 19.60 \\
plasma-cell & 0.55 & 5.61 & 56.30 +- 22.86 \\
neutrophil & 0.13 & 0.89 & 80.78 +- 41.71 \\
eosinophil & 0.10 & 0.68 & 87.97 +- 45.91 \\
\end{tabular}
\caption{Dataset statistics}
\label{tab:data_stats}
\end{table}
The background class outweighs all nuclei classes with 83.97\% of all pixels, yet it does contain cell bodies for the corresponding nuclei. Connective tissue cells include fibroblasts, endothelial cells and myocytes. Epithelial cells include both non- and neoplastic epithelial cells.

\subsection{Model Architecture}
We evaluated three different model architectures on our internal validation set (see Table~\ref{tab:preliminary_results}). Models differ only in their encoder architecture. The larger EfficientNet B7 considerably improved performance compared to EfficientNet B5. Likewise, the yet larger EfficientNet L2 yielded another considerable improvement  over B7.
%
%\begin{table}[htb!]
%\centering
%\begin{tabular}{l|lll}
%Model & mPQ+ & PQ & R2 \\ \hline
%U-Net L2 & 0.5729 & 0.6622 & 0.9069 \\
%U-Net B7 & 0.5527 & 0.6507 & 0.8802 \\
%U-Net B5 & 0.5361 & 0.6334 & 0.8701 \\
%\end{tabular}
%\caption{Validation set results for three model architectures}
%\label{tab:preliminary_results}
%\end{table}
%
\begin{table}[htb!]
\centering
\begin{tabular}{l|ll|llllll}
&&&&& PQ+ &&\\
Model & mPQ+ & R2 & neu & epi & lym & pla & eos & con\\ \hline
U-Net L2 & .572  & .906 & .429 & .641 & .713 & .573 & .451 & .630 \\
U-Net B7 & .553  & .880 & .408 & .629 & .699 & .547 & .421 & .611 \\
U-Net B5 & .536  & .870 & .376 & .622 & .694 & .530 & .396 & .599 \\
\end{tabular}
\caption{Validation set results for three model architectures}
\label{tab:preliminary_results}
\end{table}

\subsection{Class Balancing Ablation}
We conducted an ablation study to assess the impact of our importance sampling (smpl \xmark\text{ }vs. \checkmark)  and loss-weighting (lw \xmark\text{ }vs. \checkmark) methods on model performance on the Challenge metrics. Both sampling and loss weighting increase performance considerably as shown in Table~\ref{tab:ablation}. 
% By the time of submission training did not finish for the loss weighting only model, therefore results in Table~\ref{tab:ablation} will be updated for final submission.
As expected, rare cell types are  drastically affected by the absence of importance sampling and/or loss weighting, while performance on abundant cell types remains comparable.
%%%
%\begin{table}[h]
%\centering
%\begin{tabular}{cc|lll}
%sampling & loss weight. & mPQ+   & PQ     & R2     \\ \hline
%\checkmark &\checkmark & .553 & .651 & .880 \\
%\xmark & \checkmark & .405 & .586 & .661 \\
%\xmark & \xmark & .352 & .587 & .392
%\end{tabular}
%\caption{Ablation study: We optimized U-Net B7 models with different configurations to assess the impact of our loss weighting and importance %sampling methods.}
%\label{tab:ablation}
%\end{table}
\begin{table}[h]
\centering
\begin{tabular}{cc|ll|llllll}
&&&&& PQ+ &&\\
smpl & lw & mPQ+      & R2 & neu & epi & lym & pla & eos & con\\ \hline
\checkmark &\checkmark & .553 & .880 & .408 & .629 & .699 & .547 & .421 & .611 \\
\xmark & \checkmark & .405 & .661 & .153 & .572 & .551 & .391 & .265 & .496\\
\xmark & \xmark & .352 & .392 & .047 & .564 & .563 & .386 & .046 & .506
\end{tabular}
\caption{Ablation study: We optimized U-Net B7 models with different configurations to assess the impact of our loss weighting and importance sampling methods.}
\label{tab:ablation}
\end{table}
%%%

% \subsection{Model Ensemble}
% here goes the B7+L2 ensemble result

\subsection{Post-processing Ablation}
% here go the ablations on the various post-processing steps and hyperparameter searches (type-agnostic vs specific)
We ablated our post-processing pipeline with predictions from the U-Net B7 model. In table~\ref{tab:postprocessing-ablation} we present the benchmark metrics for cell-type agnostic vs. cell-type specific seed thresholds (CTS seed \xmark\text{ }vs. \checkmark) and fg thresholds (CTS fg \xmark\text{ }vs. \checkmark). Furthermore, connected components with cell-type specific minimum and maximum size thresholds for instance mask removal and solidity threshold-based instance mask removal 
is ablated by exclusion vs. inclusion into the pipeline (\xmark\text{ }vs. \checkmark).
%In our B7 model, our postprocessing, i.e., connected component analysis and small/large component removal, when performed with cell-type agnostic thresholds, yields a total 0.018 increase in mPQ+. Picking cell type specific (CTS) minimum and maximum size thresholds for instance mask removal yields another 0.0017 increase in mPQ+. Picking cell type specific foreground- and seed thresholds for watershed yields another 0.01 increase in mPQ+.  Removing instance masks by means of their solidity yields another 0.001 increase in mPQ+. 
% cell-type specific foreground and cell-type specific seed threshold, small/large component removal, solidity
\begin{table}[h]
\centering
\begin{tabular}{cccc|ll}
CTS seed & CTS fg & CC removal & solidity & mPQ+ & R2     \\ \hline
\checkmark &\checkmark & \checkmark &\checkmark & .553 & .880 \\
\xmark & \checkmark & \checkmark & \checkmark & .552 &  .866 \\
\xmark & \xmark & \checkmark & \checkmark & .547  & .848 \\
\xmark & \xmark & \xmark & \checkmark & .523 & .845 \\
\xmark & \xmark & \xmark & \xmark & .512 & .833
\end{tabular}
\caption{We ablate our post-processing pipeline and report results for fixed predictions from a U-Net B7 model}
\label{tab:postprocessing-ablation}
\end{table}
\subsection{Generalization Across Centers}
Our default training/validation split of the provided dataset does not take into account which image a sample stems from. Thus samples from the same image may be spread across training and validation set. Consequently, when using this split, it cannot be ruled out that results are, to some extent, due to overfitting to some characteristics of individual images. 

To cleanly measure the generalization ability of our B7 and L2 models, we trained them on a split that does not just split by image, but \emph{by center}, where the GlaS subset of the dataset (702 tiles) constitutes our validation set. Table~\ref{tab:center_split} lists results. 

The L2 variant improves upon B7 by a similar amount as on our default training/validation split (cf.\ Table~\ref{tab:preliminary_results}), suggesting that the (much larger) L2 variant does not overfit more to individual image characteristics than B7. On the other hand, we do observe a  decrease in performance overall, suggesting a considerable domain shift between centers.
%\begin{table}[h]
%\centering
%\begin{tabular}{l|lll}
%Model    & mPQ+   & PQ  & R2     \\ \hline
%U-Net L2 & 0.4589 & 0.6206 & 0.7113 \\
%U-Net B7 & 0.4303 & 0.6049 & 0.4395
%\end{tabular}
%\caption{Generalization study: models were trained on 4279 tiles from DigestPath, CRAG, CoNSeP and PanNuke subsets of the lizard dataset and validated on 702 tiles from GlaS subset.}
%\label{tab:center_split}
%\end{table}
%
\begin{table}[h]
\centering
\begin{tabular}{l|ll|llllll}
&&&&& PQ+ &&\\
Model & mPQ+ & R2 & neu & epi & lym & pla & eos & con\\ \hline
U-Net L2 & .458 & .711 & .214 & .615 & .638 & .424 & .350 & .514\\
U-Net B7 & .430 & .439 & .134 & .589 & .638 & .400 & .326 & .499
\end{tabular}
\caption{Generalization study: models were trained on 4279 tiles from DigestPath, CRAG, CoNSeP and PanNuke subsets of the lizard dataset and validated on 702 tiles from GlaS subset.}
\label{tab:center_split}
\end{table}
%Elias: i've put some comments in the discussion on what happens when you actually apply this on data on Bern cohort. We could include a small part here that we do this with tile and stitch? and then you can self-cite :). I have an example image for this but the results are only good if you don't go into too much detail, so we can also skip this in its entirety.

\subsection{Challenge Submission}
We submitted two different versions, one per Task, for the final test set. For \textit{Task 1}, we used an ensemble model consisting of one U-Net with EfficientNet-L2 backbone, and one U-Net with EfficientNet-B7 backbone and treat the results the same as test time augmentations from one model. We picked this ensemble as our best guess for good performance and generalization: We had observed superior performance of the L2 variant on our random train/validation split, yet given that this split is by sample and not by image, we were unable to rule out (before submission deadline) that this was due to L2 overfitting to individual images. Ensembling L2 with B7 was intended to hedge this risk. The ensemble performed very similar to the L2 variant alone on our random split (ensemble mPQ+: $0.5697$, cf.\ Table~\ref{tab:preliminary_results}). % , U-Net L2 mPQ+: $0.5729$ 

For \textit{Task 2}, we use only the U-Net with EfficientNet-B7 backbone because it had vastly outperformed competitors on the preliminary test set. We transform the output by center-cropping the masks to $224\times224$ and count the number of instances per class.

\section{Discussion}
Nuclei detection on H\&E WSI is a difficult task as ground truth labels are sometimes incorrect, nuclei shapes are modified by tissue preparation and staining, and some nuclei differentiation might be impossible on a scanned H\&E WSI at $0.5\mu pp$. Moreover, cells can be densely clustered or occur rarely, making them difficult to learn and detect correctly.
In this work, we address some of these issues by means of a 3-class loss to actively learn nuclei boundaries, as well as extensive class weighting both in the loss function as well as for the training sampling strategy. 
% Insert a part here that talks about the specific results :

%%%%%%%%%%%%%%%%%%%%%
Yet, further work in this domain is required as we still observe a number of misclassifications as well as nuclei not being detected correctly.
%TODO maybe formulate this in a way thats less offensive to the dataset creators
Moreover, the Lizard dataset~\cite{Graham2021LizardAL} aggregated some cell types (e.g. muscle, fibroblasts, endothelial cells) and excluded a number of other highly relevant cells such as macrophages and dendritic cells as well as basophils and mast cells.
% maybe (re)move this part
We also apply the proposed model to out-of-distribution data such as our own Bern colorectal cancer cohort without any adjustments. Upon visual inspection we find that most nuclei are correctly detected and even classified correctly in regions similar to the training set. However, we also find several failure modes such as transversally cut muscle where the small roundish nuclei are classified as lymphocytes, necrotic tissue or other debris where the model detects nuclei, and mitoses that are classified as neutrophils.
Furthermore, the performance of the model on WSIs could be improved by ensuring that there are no inconsistencies at the sitching boundaries when running tile-and-stitch inference~\cite{rumberger2021shift}.
%%%%%%%%%%%%%%%%%%%%%%%%%%%%%%%%%%

\section{Conclusion}

Nuclei detection, segmentation and classification on H\&E histopathology images has the potential to save time and money in diagnostics and biomarker discovery as immunohistochemistry images to infer cell types can be omitted. Moreover, large scale investigations into cellular composition can be launched as digitally scanned H\&E WSIs become readily available in most clinics. In particular, providing instance masks over detections also allows investigations into nuclei shape and orientation.

In this work, we introduce a deep learning algorithm for panoptic segmentation of nuclei in H\&E images capable of indentifying rare cell types, segmenting even in densely clustered regions and robust to stain and image quality variations.
While promising, future work should further expand data sources, differentiate between more cell types and include more tissue variations. It may also be worth investigating the same issue on higher resolution images to differentiate some cell types as intra-nucleus hematoxylin pattern could be more easily distinguishable. 
Last but not least, our generalization study strongly suggests that extending our approach by means of domain adaptation methodology is warranted to cope with considerable domain shifts between centers. 

% i switched to a proper bibtex file 
\bibliographystyle{IEEEtran}
\bibliography{IEEEabrv,bibfile}

\begin{thebibliography}{10}
\providecommand{\url}[1]{#1}
\csname url@samestyle\endcsname
\providecommand{\newblock}{\relax}
\providecommand{\bibinfo}[2]{#2}
\providecommand{\BIBentrySTDinterwordspacing}{\spaceskip=0pt\relax}
\providecommand{\BIBentryALTinterwordstretchfactor}{4}
\providecommand{\BIBentryALTinterwordspacing}{\spaceskip=\fontdimen2\font plus
\BIBentryALTinterwordstretchfactor\fontdimen3\font minus
  \fontdimen4\font\relax}
\providecommand{\BIBforeignlanguage}[2]{{%
\expandafter\ifx\csname l@#1\endcsname\relax
\typeout{** WARNING: IEEEtran.bst: No hyphenation pattern has been}%
\typeout{** loaded for the language `#1'. Using the pattern for}%
\typeout{** the default language instead.}%
\else
\language=\csname l@#1\endcsname
\fi
#2}}
\providecommand{\BIBdecl}{\relax}
\BIBdecl

\bibitem{Balkwill2012TheTM}
F.~R. Balkwill, M.~Capasso, and T.~Hagemann, ``The tumor microenvironment at a
  glance,'' \emph{Journal of Cell Science}, vol. 125, pp. 5591 -- 5596, 2012.

\bibitem{Schmidt2018CellDW}
U.~Schmidt, M.~Weigert, C.~Broaddus, and E.~W. Myers, ``Cell detection with
  star-convex polygons,'' in \emph{MICCAI}, 2018.

\bibitem{Graham2019HoverNetSS}
S.~Graham, Q.~D. Vu, S.~e~Ahmed~Raza, A.~Azam, Y.-W. Tsang, J.~T. Kwak, and
  N.~M. Rajpoot, ``Hover-net: Simultaneous segmentation and classification of
  nuclei in multi-tissue histology images,'' \emph{Medical image analysis},
  vol.~58, p. 101563, 2019.

\bibitem{Graham2021CoNICCN}
S.~Graham, M.~Jahanifar, Q.~D. Vu, G.~Hadjigeorghiou, T.~Leech, D.~R.~J. Snead,
  S.~e~Ahmed~Raza, F.~A. Minhas, and N.~M. Rajpoot, ``Conic: Colon nuclei
  identification and counting challenge 2022,'' \emph{ArXiv}, vol.
  abs/2111.14485, 2021.

\bibitem{Graham2021LizardAL}
S.~Graham, M.~Jahanifar, A.~Azam, M.~Nimir, Y.-W. Tsang, K.~C. Dodd, E.~Hero,
  H.~Sahota, A.~Tank, K.~Benes, N.~Wahab, F.~A. Minhas, S.~e~Ahmed~Raza,
  H.~Eldaly, K.~Gopalakrishnan, D.~R.~J. Snead, and N.~M. Rajpoot, ``Lizard: A
  large-scale dataset for colonic nuclear instance segmentation and
  classification,'' \emph{2021 IEEE/CVF International Conference on Computer
  Vision Workshops (ICCVW)}, pp. 684--693, 2021.

\bibitem{Araslanov2021SelfsupervisedAC}
N.~Araslanov and S.~Roth, ``Self-supervised augmentation consistency for
  adapting semantic segmentation,'' \emph{2021 IEEE/CVF Conference on Computer
  Vision and Pattern Recognition (CVPR)}, pp. 15\,379--15\,389, 2021.

\bibitem{Hirsch2020AnAT}
P.~Hirsch and D.~Kainmueller, ``An auxiliary task for learning nuclei
  segmentation in 3d microscopy images,'' \emph{ArXiv}, vol. abs/2002.02857,
  2020.

\bibitem{Caicedo2019EvaluationOD}
J.~C. Caicedo, J.~Roth, A.~Goodman, T.~Becker, K.~W. Karhohs, M.~Broisin,
  C.~Molnar, C.~McQuin, S.~Singh, F.~J. Theis, and A.~E. Carpenter,
  ``Evaluation of deep learning strategies for nucleus segmentation in
  fluorescence images,'' \emph{Cytometry}, vol.~95, pp. 952 -- 965, 2019.

\bibitem{Tan2019EfficientNetRM}
M.~Tan and Q.~V. Le, ``Efficientnet: Rethinking model scaling for convolutional
  neural networks,'' \emph{ArXiv}, vol. abs/1905.11946, 2019.

\bibitem{Xie2020SelfTrainingWN}
Q.~Xie, E.~H. Hovy, M.-T. Luong, and Q.~V. Le, ``Self-training with noisy
  student improves imagenet classification,'' \emph{2020 IEEE/CVF Conference on
  Computer Vision and Pattern Recognition (CVPR)}, pp. 10\,684--10\,695, 2020.

\bibitem{Ronneberger2015UNetCN}
O.~Ronneberger, P.~Fischer, and T.~Brox, ``U-net: Convolutional networks for
  biomedical image segmentation,'' in \emph{MICCAI}, 2015.

\bibitem{Ruifrok2001QuantificationOH}
A.~C. Ruifrok and D.~A. Johnston, ``Quantification of histochemical staining by
  color deconvolution.'' \emph{Analytical and quantitative cytology and
  histology}, vol. 23 4, pp. 291--9, 2001.

\bibitem{Loshchilov2019DecoupledWD}
I.~Loshchilov and F.~Hutter, ``Decoupled weight decay regularization,'' in
  \emph{ICLR}, 2019.

\bibitem{Gal2016DropoutAA}
Y.~Gal and Z.~Ghahramani, ``Dropout as a bayesian approximation: Representing
  model uncertainty in deep learning,'' \emph{ArXiv}, vol. abs/1506.02142,
  2016.

\bibitem{rumberger2020probabilistic}
J.~L. Rumberger, L.~Mais, and D.~Kainmueller, ``Probabilistic deep learning for
  instance segmentation,'' in \emph{European Conference on Computer
  Vision}.\hskip 1em plus 0.5em minus 0.4em\relax Springer, 2020, pp. 445--457.

\bibitem{carpenter2006cellprofiler}
A.~E. Carpenter, T.~R. Jones, M.~R. Lamprecht, C.~Clarke, I.~H. Kang,
  O.~Friman, D.~A. Guertin, J.~H. Chang, R.~A. Lindquist, J.~Moffat
  \emph{et~al.}, ``Cellprofiler: image analysis software for identifying and
  quantifying cell phenotypes,'' \emph{Genome biology}, vol.~7, no.~10, pp.
  1--11, 2006.

\bibitem{rumberger2021shift}
J.~L. Rumberger, X.~Yu, P.~Hirsch, M.~Dohmen, V.~E. Guarino, A.~Mokarian,
  L.~Mais, J.~Funke, and D.~Kainmueller, ``How shift equivariance impacts
  metric learning for instance segmentation,'' in \emph{Proceedings of the
  IEEE/CVF International Conference on Computer Vision}, 2021, pp. 7128--7136.

\end{thebibliography}

\vspace{12pt}
\end{document}